\documentclass[11pt]{article}
\usepackage{graphics}
\usepackage{psfrag}
\usepackage{epsfig}
\usepackage{epsf}
\usepackage{float}
\usepackage{amssymb,stmaryrd,latexsym}
\newcommand{\beq}{\begin{equation}}

\newcommand{\eeq}{\end{equation}}
\newcommand{\beqa}{\begin{eqnarray}}
\newcommand{\eeqa}{\end{eqnarray}}

\newcommand{\no}{\nonumber}

\begin{document}

\title{\vspace{-1.2cm}
\hfill {\tiny FZJ--IKP(TH)--2008--09, HISKP-TH-08-09}\\
\vspace{1cm}%
Subleading contributions to the width of the {$D_{s0}^*(2317)$}}

\author{Feng-Kun Guo$^{\rm a}$\footnote{{\it Email address:} f.k.guo@fz-juelich.de},
Christoph Hanhart$^{\rm a}$\footnote{{\it Email address:}
c.hanhart@fz-juelich.de}, Siegfried Krewald$^{\rm a}$\footnote{{\it
Email address:} s.krewald@fz-juelich.de} $\,\,$
and Ulf-G. Mei{\ss}ner$^{\rm a, b}$\footnote{{\it Email address:} meissner@itkp.uni-bonn.de}%
\\[2mm]
{\small $^{\rm a}${\it Institut f\"{u}r Kernphysik,
Forschungszentrum
J\"{u}lich, D--52425 J\"{u}lich, Germany}} \\[0mm]
{\small $^{\rm b}${\it Helmholtz-Institut f\"ur Strahlen- und
Kernphysik
(Theorie), Universit\"at Bonn,}}\\
{\small {\it Nu{\ss}allee 14-16, D--53115 Bonn, Germany}}}

\date{}

\maketitle

\begin{abstract}
  We construct the effective chiral Lagrangian involving the
  $D$--mesons and Goldstone bosons at next-to-leading order taking
  into account strong as well as electromagnetic interactions. This
  allows us to disentangle --- to leading order in isospin violation
  --- the electromagnetic and the strong contribution to the
  $D$--meson mass differences. In addition, we also apply the
  interaction to the decay $D_{s0}^*(2317)\to D_s\pi^0$ under the
  assumption that the $D_{s0}^*(2317)$ is a hadronic molecule. We find
  $(180\pm 110)$ keV for the decay width $\Gamma ({D_{s0}^*(2317)\to
    D_s\pi^0})$ --- consistent with currently existing experimental
  constraints as well as previous theoretical investigations.  The
  result provides further evidence that this decay width can serve as
  a criterion for testing the nature of the $D_{s0}^*(2317)$.
\end{abstract}

\vspace{1cm}

\section{Introduction}

In recent years, many heavy mesons with open or hidden charm were
discovered, which contribute to the revival of hadron spectroscopy
(for recent reviews, see \cite{charmreview}). One outstanding example
among them is the $D_{s0}^*(2317)$ discovered by the BABAR
Collaboration in the $D_s\pi$ final state~\cite{Aubert:2003fg}. The
measured mass of the $D_{s0}^*(2317)$ which is
$2317.8\pm0.6$~MeV~\cite{Yao:2006px} is much lower than that predicted
in many quark models.  An appealing alternative is that the
$D_{s0}^*(2317)$ is a hadronic molecule, which means that it owes its
existence to meson--meson dynamics~\cite{Barnes:2003dj,eef}.  In this
work we exploit further this idea.  Note, in Refs.~\cite{mehen} it was
argued that a molecular interpretation of the $D_{s0}^*(2317)$ (and
its vector counter part) is at variance with heavy quark effective
field theory. However, this conclusion is based on the assumption that
the decay of a hadronic molecule is proportional to the molecular wave
function at the origin --- in Ref.~\cite{f0gg} it is shown that this
assumption is not justified for the decay of hadronic molecules.

The mass alone is not a signal for a molecule, as stressed, e.g., in
Ref.~\cite{Guo:2007up}. A consistent treatment of the mass and various
decays is required --- see also Ref.~\cite{evidence}. The difference
from the molecular state will presumably be revealed in the decay
pattern into various channels. However, no branching ratio of the
$D_{s0}^*(2317)$ has been reported accurately. The only experimental
constraints are upper limits for the ratios of some other decay
channels to the $D_{s0}^*(2317)^+\to D_s^+\pi^0$. For
instance~\cite{Yao:2006px},
\begin{equation}
\label{eq:exp} {\Gamma({D_{s0}^*(2317)^+\to D_s^*(2112)^+\gamma})
\over\Gamma ({D_{s0}^*(2317)^+\to D_s^+\pi^0})} < 0.059.
\end{equation}

The $D_{s0}^*(2317)$ can be dynamically generated from Goldstone
boson--$D$-meson scattering as a hadronic molecule using unitarized
amplitudes from chiral pertrubation
theory~\cite{Hchua,lutz04npa,Guo:2006fu,Lutz:2007sk}. The width of the
isospin violating decay $D_{s0}^*(2317)^+\to D_s^+\pi^0$ was estimated
to be about $8.7\,$keV in Ref.~\cite{Guo:2006fu} by considering only
the $\pi^0$--$\eta$ mixing in the final state. However, Faessler {\it
  et al}. pointed out that the mass differences between neutral and
charged kaons and $D$--mesons give an important contribution
\cite{Faessler:2007}, which was confirmed later in
Ref.~\cite{Lutz:2007sk}.  In Refs.~\cite{lutz04npa,Lutz:2007sk} also
subleading operators were studied. Note that isospin symmetry
violation in hadronic physics has two different sources: one
originates from the mass difference of the light $u$ and $d$ quarks,
and the other one stems from the electromagnetic (e.m.) interaction.
While the $\pi^0$--$\eta$ mixing and part of the meson mass
differences account for the former one, the effect of the latter on
the decay $D_{s0}^*(2317)^+\to D_s^+\pi^0$ will be investigated here
for the first time.

The first step is to construct the interaction Lagrangian to
next--to--leading order in the chiral expansion.  Based on this we
can disentangle the e.m. and the strong contribution to the
$D$--meson mass difference.  As will be demonstrated, the Lagrangian
also links these mass differences directly to the isospin--violating
Goldstone boson--$D$-meson scattering amplitudes, in full analogy to
the case of $\pi N$ scattering and the proton--neutron mass
difference~\cite{wein,Meissner:1997ii}. We can therefore calculate
the decay width of the $D_{s0}^*(2317)^+\to D_s^+\pi^0$ --- within
the molecular picture utilizing the chiral Lagrangian up to the
next-to-leading order (NLO) ${\mathcal O}(p^2)$, where $p$ denotes a small
parameter --- with a small number of free parameters. Note, in this
paper for the first time both the strong and the e.m. contributions
to the decay are incorporated systematically.

\section{Lagrangians at next-to-leading order}

The scattering between the Goldstone bosons and $D$--mesons is
similar to the case for pion-nucleon scattering (for reviews, see
Refs.~\cite{Bernard:1995dp,Bernard:2007zu}) because the $D$--mesons
have heavy masses which do not vanish in the chiral
limit\footnote{In this work we consider the SU(3) chiral limit,
$m_u, \, m_d,\, m_s\to 0$.}. We count the $D$--meson masses ($\sim
1.9$~GeV) as order ${\mathcal O}(\Lambda_{\chi}) \sim {\mathcal O}(p^0)$ where
$\Lambda_\chi\simeq 1$~GeV. Hence the leading order terms in the
chiral Lagrangian are of ${\mathcal O}(p)$, and the NLO terms are of ${\mathcal O}(p^2)$.

The leading order Lagrangian is just the kinetic energy term of the
heavy mesons~\cite{hclo}
\begin{equation}
{\cal L}^{(1)} = {\cal D}_{\mu}D{\cal D}^{\mu}D^{\dag}-m_D^2
DD^{\dag}
\end{equation}
with $D=(D^0,D^+,D_s^+)$ denoting the $D$--mesons, and the covariant
derivative being
\begin{eqnarray}
{\cal D}_{\mu} &\!\!=&\!\! \partial_{\mu}+\Gamma_{\mu}, \nonumber\\
\Gamma_{\mu} &\!\!=&\!\!
{1\over2}\left(u^{\dag}\partial_{\mu}u+u\partial_{\mu}u^{\dag}\right),
\end{eqnarray}
where
\beqa%
U = \exp \left( {\sqrt{2}i\phi \over F_\pi}\right),\quad u^2=U.
\eeqa%
The Goldstone boson fields are collected in the matrix
\beqa%
\label{eq:phi}
 \phi =
  \left(
    \begin{array}{c c c}
 \frac{1}{\sqrt{2}}\pi^0+\frac{1}{\sqrt{6}} \eta & \pi^+ & K^+ \\
\pi^- & - \frac{1}{\sqrt{2}}\pi^0+\frac{1}{\sqrt{6}} \eta & K^0 \\
K^- & \bar K ^0 & -\frac{2}{\sqrt{6}} \eta \\
    \end{array}
\right) .
\eeqa%

We now consider the NLO chiral Lagrangian describing the interactions of
the pseudoscalar charm mesons with the Goldstone bosons. Considering
the heavy mesons as  matter fields, similarly to the
pion-nucleon sector~\cite{Meissner:1997ii}, the strong part is
\beqa%
\label{eq:L2str} {\cal L}^{(2)}_{\rm str.} &\!\!=&\!\! D \left(
-h_0\langle\chi_+\rangle - h_1\tilde{\chi}_+ + h_2\left\langle
u_{\mu}u^{\mu} \right\rangle - h_3u_{\mu}u^{\mu}
\right) {\bar D} \no\\
&\!\!&\!\! + {\cal D}_{\mu}D \left( h_4\langle u^{\mu}u^{\nu}\rangle
- h_5 \{u^{\mu},u^{\nu}\} - h_6 [u^{\mu},u^{\nu}] \right) {\cal
D}_{\nu}{\bar D} .
\eeqa%
The chiral symmetry breaking terms, i.e. $h_0$ and $h_1$ terms, have
been introduced before~\cite{Cheng:1993kp,lutz04npa}. The $h_2$ and
$h_3$ terms were introduced in Ref.~\cite{lutz04npa}.
We stress that the contributions of the $h_5$ and $h_3$ terms to $s$--wave
amplitudes differ only to order ${\mathcal O}(p/m_D)$. However, we still
keep them in our covariant formalism for in this way we have an additional tool
to estimate the theoretical uncertainty --- see section~\ref{sec:results}.
 The
electromagnetic part is
\beq%
\label{eq:L2em} {\cal L}^{(2)}_{\rm e.m.} = F_{\pi}^2 D \left[ g_0
\left({Q_+^2-Q_-^2}\right) + g_1
\left\langle{Q_+^2-Q_-^2}\right\rangle + g_2{{Q_+}}\left\langle
Q_+\right\rangle + g_3\left\langle Q_+\right\rangle^2 \right] {\bar
D} \ ,
\eeq%
 where
\beqa
\chi_+ &\!\!=&\!\! u^\dagger \chi u^\dagger + u\chi u ,\nonumber\\
\tilde{\chi}_+ &\!\!=&\!\! \chi_+ -
{1\over3}\left\langle\chi_+\right\rangle,\nonumber\\
u_{\mu} &\!\!=&\!\! iu^{\dag}{\cal D}_{\mu}Uu^{\dag}, \nonumber\\
Q_\pm &\!\!=&\!\! \frac12\left( u^\dagger Q u \pm uQu^\dagger
\right).
\eeqa%
The quark mass matrix and the $D$--meson charge matrix are
diagonal
\begin{equation}
\chi = 2B\cdot {\rm diag}\left\{m_u,m_d,m_s\right\}, \quad Q =
e\cdot {\rm diag}\left\{0,1,1\right\}~,
\end{equation}
in terms of $B= |\langle 0 |\bar q q |0\rangle|/F_{\pi}^2 $ and the
elementary charge $e$. Further, $F_{\pi}$ is the pion decay constant.
The unknown coefficients $h_i~(i=0,1,...,6)$ and $g_i~(i=0,1,2,3)$ in
Eqs.~(\ref{eq:L2str},\ref{eq:L2em}) are the so-called low energy
constants (LECs). As we will show in the next section, $h_1$ and a
linear combination of $g_0$ and $g_2$, namely $g_0+2g_2$, can be
determined from the mass differences among the $D$--mesons.  Since
$\langle Q_+\rangle=2e$, the $g_3$ term in ${\cal L}^{(2)}_{\rm e.m.}$
only gives an overall e.m. mass shift of the $D$--mesons, and hence
can be absorbed in the bare masses. The $g_1$ term contains only
isospin symmetric e.m. interaction and is irrelevant here.  Terms with
one more flavor trace in the strong interaction Lagrangian are
suppressed in the large $N_C$ limit of QCD~\cite{Manohar:1998xv}. We
therefore follow Ref.~\cite{Lutz:2007sk} and drop the $h_0$, $h_2$ and
$h_4$. Formally the $h_6$ term is of ${\mathcal O}(p^2)$. However, due
to the commutator structure, it is suppressed by one order, see
Appendix~\ref{app:h6}. We are therefore left with only two free,
active parameters, both isospin conserving, namely $h_3$ and
$h_5$. We will investigate their effect on the isospin violating decay
of the $D_{s0}^*$ below.

One should note that the $\pi^0$ and $\eta$ in Eq.~(\ref{eq:phi})
are not mass eigenstates because of the  $\pi^0$--$\eta$ mixing. The mass
eigenstates are defined as
\begin{eqnarray}
\tilde{\pi}^0 &\!\!=&\!\!
\pi^0\cos\epsilon_{\pi^0\eta}+\eta\sin\epsilon_{\pi^0\eta},
\nonumber\\
\tilde{\eta} &\!\!=&\!\! -\pi^0\sin\epsilon_{\pi^0\eta}
+\eta\cos\epsilon_{\pi^0\eta},
\end{eqnarray}
where $\epsilon_{\pi^0\eta}$ is the well-known $\pi^0$--$\eta$
mixing angle, which reads to leading order
\beq%
\epsilon_{\pi^0\eta}
= \frac{\sqrt{3}}{4}\frac{m_d-m_u}{m_s-\hat m}
\eeq%
with $\hat m=(m_u+m_d)/2$ the average mass of the $u$ and $d$
quarks.

\section{$D$--meson mass differences}

The terms which can contribute to the mass differences among the
$D^+,~D^0$ and $D_s^+$ mesons come from the Lagrangian of order
${\mathcal O}(p^2)$, Eqs.~(\ref{eq:L2str},\ref{eq:L2em}). Only three among all
the terms  contribute.
 We find
\beqa%
m_{D^0}^2-m_{D^+}^2 &=& \bar h \lambda + \bar g, \nonumber\\
m_{D^+}^2-m_{D^+_s}^2 &=& \bar h\left(1-\frac{ \lambda}{2}\right),
\label{parafix}
\eeqa%
where we define
\beq%
\bar h = 4Bh_1(m_s-\hat m), \quad \bar g = F_{\pi}^2 e^2(g_0+2g_2).
\label{bardefs}
\eeq%
The strength of
isospin violation due to quark mass effects is encoded in the
parameter $\lambda$ that is connected to $\epsilon_{\pi^0\eta}$
through
\beq \lambda = \frac{m_d-m_u}{m_s-\hat m}
=\frac{4}{\sqrt{3}}\epsilon_{\pi^0\eta}. \eeq
 A recent analysis of
$\rho$--$\omega$ mixing in chiral perturbation theory gives
$1/\lambda=42\pm4$~\cite{Kucukarslan:2006wk}, correspondingly we have
$\lambda=0.024\pm 0.002$. Using the masses for the
$D$--mesons~\cite{Yao:2006px}, $m_{D^0}=1864.84\pm 0.17$ MeV,
$m_{D^+}=1869.62\pm 0.20$ MeV, and $m_{D_s^+}=1968.49\pm 0.34$ MeV,
we find
\beq%
\bar h = (384.1\pm 2.5)\times 10^3~{\rm MeV}^2, \quad \bar g =
(4\pm1)\times 10^3~{\rm MeV}^2 ,
\eeq%
where the largest uncertainty comes from the masses of the
$D$--mesons. Note that here only the uncertainties from the
experimental inputs are considered. For a discussion of the
theoretical uncertainty, see Section~\ref{sec:results}. Then the
dimensionless LECs $h_1$ and $g_0+2g_2$ can be determined as
\beq%
\label{eq:lec} h_1 = 0.42\pm0.00 , \quad g_0+2g_2 = 11\pm 3,
\eeq%
where we use
$B(m_s-\hat{m})=\left(M_{K^0}^2+M_{K^+}^2\right)/2-M_{\pi^0}^2$. The
basic assumption in setting up an effective field theory is the
naturalness of the low energy constants --- especially the
dimensionless coefficient $h_1$ should be of order one. This is
indeed the case for we find $h_1 = 0.42$. A naturalness estimate for
$g_0$ and $g_2$ comes from requiring that the contribution
of the corresponding operators to the $D$--meson mass shift
 should be of the order
of a typical virtual
photon loop~\cite{Fettes:2000vm}, thus
\begin{equation}
e^2F_{\pi}^2g_i\sim \left({e\over 4\pi}\right)^2m_D^2.
\end{equation}
This leads to $g_i\sim 4$ as a natural estimate of the order of
magnitude, compatible with  the value determined for $g_0+2g_2$.

The parameter $\bar h$, fixed from the amount of $SU(3)$ violation
encoded in the mass difference between the $D_s^+$ and the $D^+$
(see Eq.~(\ref{parafix})), controlls also the strong part of the $D^0$
and $D^+$ mass difference. Therefore, the electromagnetic
contribution to this mass difference, which is given by the ${\bar
g}$ term, can be extracted from data. This is different to the case of,
e.g., nucleons, where the operator structure is more complicated. We
therefore get
\begin{eqnarray}
\left(m_{D^+}-m_{D^0}\right)_{\rm str.} &\!\!=&\!\!
(2.5\pm0.2)~{\rm MeV}, \\
\left(m_{D^+}-m_{D^0}\right)_{\rm e.m.} &\!\!=&\!\!
(2.3\pm0.6)~{\rm MeV},
\end{eqnarray}
where the first equation refers to the strong contribution to the
mass difference and the second to its e.m. counterpart. Note,
contrary to what is common for nucleons as well as kaons, here the
electromagnetic  and the strong effects enter with the same sign.
This is a direct consequence of the different quark content of the
states.

The strong and e.m. mass differences of $m_{D^+}-m_{D^0}$ are
consistent with those determined long time ago by Gasser and
Leutwyler using a simple quark model ansatz~\cite{Gasser:1982ap},
which are $3.3\pm0.9$~MeV and $1.7\pm0.5$~MeV, respectively.

\section{Width of the {\boldmath$D_s^*(2317)$}}

For studying isospin violating decays, it is better to work in the
particle basis. For the scalar charm-strange sector, there are four
channels involving a $D$--meson and a Goldstone boson: $D^0K^+$,
$D^+K^0$, $D_s^+\eta$ and $D_s^+\pi^0$. One can expect that the
isospin violating contributions to the mass of the $D_{s0}^*(2317)$
is negligible, therefore we only consider them for calculating the
isospin violating decay width. The strong contribution to this decay
is given in terms of the known $\pi^0$--$\eta$ mixing angle and the
meson mass differences.
The only non-vanishing e.m. contribution is from
the transition $D^0K^+\to D_s^+\pi^0$
\beq%
V_{D^0K^+\to D_s^+\pi^0}^{\rm e.m.} = -{\sqrt{2}\over
8}(g_0+2g_2)e^2.
\eeq%
Especially, the amplitude for $D_s^+\eta\to D_s^+\pi^0$
vanishes.
The linear combination of LECs $(g_0+2g_2)$
has been determined from the $D$--meson mass differences in the previous section.
Thus, all relevant isospin violating interactions are fixed from data.

\subsection{Unitarization of the scattering amplitudes  at next-to-leading order}

A unitarization procedure was proposed in Ref.~\cite{Oller:2000fj}
which can be used for any finite order in the chiral expansion. Similar
to the case for pion-nucleon scattering, in our case, up to NLO
there is no loop contribution. Hence we obtain the following
$T$--matrix equation after matching to the chiral expansion at NLO
\begin{equation}
T(s)=V(s)\left[1-G(s)\cdot V(s)\right]^{-1},
\end{equation}
with $V(s)=V_{\rm LO}(s)+V_{\rm NLO}(s)$ the sum of the $S$--wave
scattering amplitudes of the LO and NLO orders~\cite{Oller:2000fj}.
$G(s)$ is a diagonal matrix with the diagonal element given by the
two-meson loop integral~\cite{Oller:2000fj,Oller:1998zr}
\begin{eqnarray}
\label{eq:G} G(s)_{ii} &\!=&\! i \int{d^4q\over (2\pi)^4} {1\over
\left(q^2-m_1^2+i\epsilon\right)\left[(P-q)^2-m_2^2+i\epsilon\right]}
\nonumber\\
&\!=&\! \frac{1}{16\pi^2}\left\{a(\mu)+\ln{\frac{m_2^2}{\mu^2}} +
\frac{m_1^2-m_2^2+s}{2s}\ln{\frac{m_1^2}{m_2^2}}
+\frac{\sigma}{2s}\left[\ln({s-m_1^2+m_2^2+\sigma})\right.\right.\nonumber\\
&\!&\! \left.\left. -\ln({-s+m_1^2-m_2^2+\sigma}) +
\ln({s+m_1^2-m_2^2+\sigma})-\ln({-s-m_1^2+m_2^2+\sigma}) \right]
\right\},
\end{eqnarray}
where $a(\mu)$ is the subtraction constant, $\mu$ denotes the scale
of the dimensional regularization, and
$\sigma=\left\{[s-(m_1+m_2)^2][s-(m_1-m_2)^2]\right\}^{1/2}$. In our analysis we
use the somewhat lengthy expression given above, for it allows for a
straightforward analytic continuation into the complex plain,
contrary to more compact representations that are
applicable in a particular parameter space only.

The subtraction constant $a(\mu)$ is determined by fitting the mass
of the $D_{s0}^*(2317)$ using  the LO Lagrangian. It turns out to be
$a(\mu=1~{\rm GeV})=-1.846$ for reproducing
$m_{D_{s0}^*}=2317.8$~MeV~\cite{Yao:2006px}. Probably by accident,
it coincides exactly with that obtained from matching the value of
the loop function at the threshold of the $D$ and $K$ calculated by
Eq.~(\ref{eq:G}) with that calculated by using a 3-momentum cut-off
$q_{\rm max}=m_{\rho}$~\cite{Guo:2006fu}.

\subsection{Results}

\label{sec:results}

In calculations, we take physical values of all the meson masses and
the pion decay constant, as listed in the
following~\cite{Yao:2006px}:
\begin{eqnarray}
&& F_{\pi}=92.42~{\rm MeV},\quad M_{\pi^0}=134.98~{\rm MeV},\quad
M_{\pi^+}=139.57~{\rm MeV},
\nonumber\\
&& M_{K^0}=497.65\pm0.02~{\rm MeV},\quad
M_{K^+}=493.68\pm0.02~{\rm MeV},\nonumber\\
&& m_{D^0}=1864.84\pm0.17~{\rm MeV},\quad
m_{D^+}=1869.62\pm0.20~{\rm MeV},\nonumber\\
&& m_{D_s^+}=1968.49\pm0.34~{\rm MeV},\quad
M_{\eta}=547.51\pm0.18~{\rm MeV}.
\end{eqnarray}

After unitarization,
the hadronic molecule $D_{s0}^*(2317)$ appears as a pole
 in the second
Riemann sheet at $\left(m_{D_{s0}^*}-i\Gamma(D_{s0}^*\to
D_s\pi^0)/2\right)$. Denoting the three-momentum of one particle in
the center-of-mass frame of
 channel $i$ by $k_{i}$, the second Riemann sheet is
specified by ${\rm Im}\,k_{D_s^+\pi^0}<0$ and ${\rm Im}\,k_{i}>0$
($i=D^0K^+,~D^+K^0,~D_s^+\eta$). First, let us focus on the results
considering the LO amplitudes only.
There are two different kinds of
contributions to the isospin violating decay width at leading order.
One is from the $\pi^0$--$\eta$ mixing
 and the other one is from
the mass differences between charged and neutral kaons and
$D$--mesons, which predominantly enters through
an isospin violating contribution to the loop function
of Eq.~(\ref{eq:G}).
However,
$\epsilon_{\pi^0\eta}=1/\sqrt{3}B(m_d-m_u)/\left(M_{\eta}^2-M_{\pi^0}^2\right)$
is suppressed by $M_\pi^2/M_{\eta}^2$ --- in SU(2) chiral perturbation
theory, where the strange quark is also viewed as heavy, this operator
appears only at NNLO, even an order below those given in
Eqs.~(\ref{eq:L2str}) and (\ref{eq:L2em}). Therefore, one can expect
that the mass differences give a larger contribution. The results
confirm this expectation as shown in the second and third column in
Table~\ref{tab:width}, corresponding to the widths considering only
the $\pi^0$--$\eta$ mixing and meson mass differences, respectively.
Furthermore, similar to Refs.~\cite{Faessler:2007,Lutz:2007sk}, in our
calculation the interference between these two kinds of contributions
are constructive, giving rise to a width of about 150~keV --- see the
first column of Table~\ref{tab:width}.  However, when it comes to a
quantitative comparison, the result for the width to leading order of
Ref.~\cite{Lutz:2007sk} is smaller by a factor of two--- a direct
comparison with the more phenomenological work of
Ref.~\cite{Faessler:2007} is not possible. The difference can be
traced to differences in the input parameters and a different method
to fix the subtraction constant $a(\mu)$ of Eq.~(\ref{eq:G}).  Those
differences should be of higher order. Thus the spread in the reported
results calls for a calculation to next--to--leading order in the
chiral expansion, c.f. Ref.~\cite{Lutz:2007sk}, together with an
analysis of the uncertainties.

\begin{table}[t]
\begin{center}
\begin{tabular}{|ccc|}\hline\hline
 LO & $\pi^0$--$\eta$ mixing & Mass differences \\ \hline%
149.4 & 15.0 & 69.7 \\ \hline\hline%
\end{tabular}
\smallskip
\caption{\label{tab:width}Decay widths of the $D_{s0}^*(2317)\to
D_s\pi^0$ with LO amplitudes. All units of the decay widths are in
keV.}

\end{center}
\end{table}

\begin{figure}[t]
\begin{center}\vspace*{0.0cm}
\includegraphics[width=0.67\textwidth]{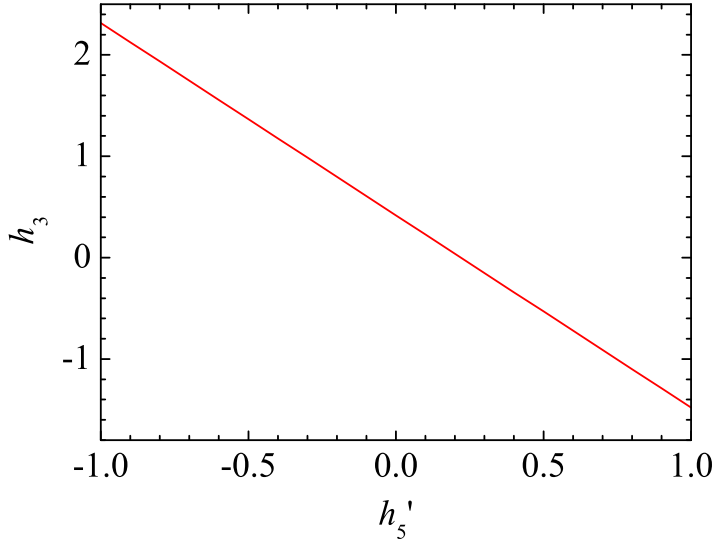}%
\vglue -0.65cm\caption{\label{fig:h3} Resulting $h_3$ from fitting
the mass of the $D_{s0}^*(2317)$ to 2317.8~MeV in the isospin
symmetric case for given $h_5'$.}
\end{center}
\end{figure}
 We
take $[-1,1]$ as a natural range for the dimensionless parameter
$h_5'\equiv h_5/m_{D^0}^2$ --- note that for $h_5'=\pm1$ the
contribution of the $h_5$ term to the $D^0K^+\to D^0K^+$ scattering
amplitude is of the same order as the leading one. The subtraction
constant is kept fixed to $a(1~{\rm GeV})=-1.846$. The value of
$h_3$ is then determined from fitting the pole position in the
second Riemann sheet to the mass of the scalar charm meson
$m_{D_{s0}^*}=2317.8\pm0.6$~MeV in the isospin symmetric case. For
each value of $h_5'$, there is a corresponding $h_3$, as shown in
Fig.~\ref{fig:h3}.

\begin{figure}[htb]
\begin{center}\vspace*{0.0cm}
\includegraphics[width=0.67\textwidth]{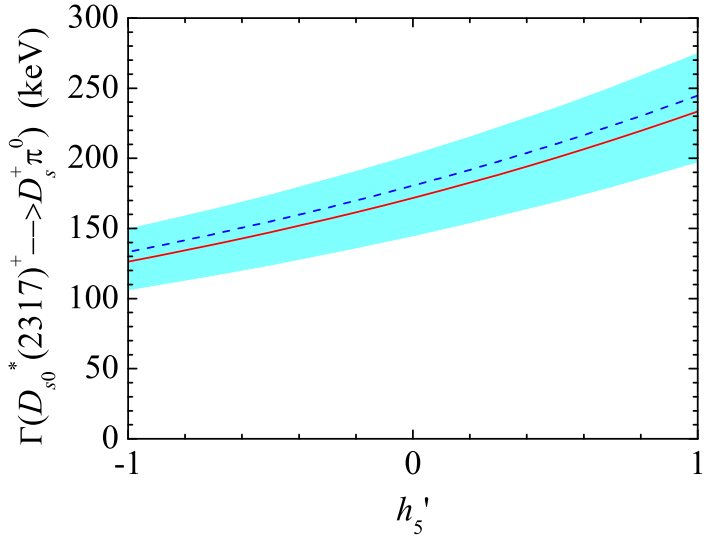}%
\vglue -0.5cm\caption{\label{fig:wid_h5}
$\Gamma\left(D_{s0}^*(2317)^+\to D_s^+\pi^0\right)$ as a function of
$h_5'$. The solid curve results from choosing the central value for
$\lambda$, while the grey band is found by varying $\lambda$ within
its allowed bounds and taking into account theoretical uncertainty.
The dashed curve refers to neglecting the e.m. term in the decay
(let $g_0=-2g_2$).}
\end{center}
\end{figure}

For each value of $h_5'$
the resulting widths are plotted in
Fig.~\ref{fig:wid_h5}.
In the figure, the solid line is the result for producing the mass
of the $D_{s0}^*(2317)$ at 2317.8 MeV with both  the strong and e.m.
isospin violating contributions and using central values for all the
parameters. The dashed curve represents the result without e.m.
contributions. The uncertainties from the experimental inputs
are reflected in the shaded band.
To estimate the theoretical uncertainty observe that, as
stressed above, the effect of the $h_3$ and the $h_5$ term
differ at ${\mathcal O}(p/m_D)$. Therefore, since one of
the two is fixed already from the mass of the $D_{s0}$,
the dependence of the width on $h_5$ is a measure of
(some) NNLO effects. We may read off the figure directly
a spread of about 50 keV around the central value of 180 keV
induced by the variation of $h_5$. To be on the safe side
we take as the theoretical uncertainty of our calculation twice
this spread. Another way to estimate the theoretical uncertainty
is to use $2(M_K/\Lambda_{\chi})^2\simeq50\%$, since we include
contributions to the amplitude up to next--to--leading order in
$(M_K/\Lambda_{\chi})$ in our calculation. The factor of 2 appears
since the width is proportional to the square of the amplitude.
It is reassuring that both methods lead to essentially similar numbers
for the uncertainty. Our final result therefore reads
\begin{equation}
\Gamma\left(D_{s0}^*(2317)^+\to D_s^+\pi^0\right) = \left(
180\pm 40 \pm 100 \right)~{\rm keV},
\end{equation}
where the first error is from experimental inputs and the second
reflects the theoretical uncertainty.
Amongst the former uncertainties,
the largest ones are from the uncertainties of the $D$--meson masses
and the $\pi^0$--$\eta$ mixing parameter $\lambda=0.024\pm0.002$.
Schematically, let us consider the case corresponding to $h_5'=1$.
The central result for the  width is 233 keV.
 When the central values of all
the meson masses are taken, the width can change from 219~keV for
$\lambda=0.022$ to 248~keV for $\lambda=0.026$. When we take the
central value $\lambda=0.024$ and the central values of the masses
of all the mesons except $D^0$ and $D^+$, the width can change from
233~keV for taking the central values of $m_{D^0}$ and $m_{D^+}$ to
249~keV for taking $m_{D^0}=1864.65$~MeV and $m_{D^+}=1869.82$~MeV.
Among all the others, the uncertainties caused by $g_0+2g_2$, see
Eq.~(\ref{eq:lec}), and the masses of kaons are the largest, and
they amount to an uncertainty of 3~keV and 1~keV at most,
respectively. All other terms give negligible contributions to the
uncertainty.

Within the molecular picture, different calculations gave the width of
the radiative decay $D_{s0}^*(2317)\to D_s^*\gamma$ in the range from
$1-6$~keV~\cite{Gamermann:2007bm,Faessler:2007,Lutz:2007sk}.
Combining with the experimental result in Eq.~(\ref{eq:exp}), the
lower limit for the width of the decay $D_{s0}^*(2317)\to D_s\pi^0$ is
of the order of 100~keV. Our result is compatible with this extracted
lower limit.


\section{Summary and Outlook}

In this paper we investigate the isospin violating decay
$D_{s0}^*(2317)\to D_s\pi$ up to NLO in the chiral expansion, assuming
that the $D_{s0}^*(2317)$ is a hadronic molecule. We take into account
electromagnetic contributions systematically for the first time. Up to
order ${\mathcal O}(p^2)$, we obtain both the strong and e.m. mass
differences of the $D$--mesons. We confirm that the mass differences
between charged and neutral mesons in the same isospin multiplets play
a significant role in the decay width. The decay width of the
$D_{s0}^*(2317)\to D_s\pi$ calculated to next--to--leading order is
found to be $180\pm 110$ keV, where the uncertainties are added in
quadrature. The uncertainty is dominated by the theoretical
one.  The resulting width is consistent with the present
experimental constraint.

Our results for the hadronic decay width of the $D_{s0}^*(2317)$
within uncertainty are consistent with previous analyses of the
$D_{s0}^*(2317)$ within a molecular picture considering both the
$\pi^0$--$\eta$ mixing and the meson mass differences:
Ref.~\cite{Faessler:2007} gives $79.3\pm 32.6$ keV, while
Ref.~\cite{Lutz:2007sk} gives $76$ keV ($140$ keV) as the result at
leading (next--to--leading) order.
Assigning the $D_{s0}^*(2317)$ to be a $c{\bar s}$ meson, its
hadronic width was estimated within quark models with typical values
of the order of 10~keV~\cite{godfrey}, although larger values were
reported --- see collection of results in table II of
Ref.~\cite{Faessler:2007}, which is consistent with the analysis
utilizing heavy quark effective field theory~\cite{mehen}. Within
the tetraquark picture a similar  width as in the $c{\bar s}$ picture
was found~\cite{tetra}, which is about one order of magnitude
smaller than our predictions in the molecular picture. We therefore
conclude that the decay width of the $D_{s0}^*(2317)\to D_s\pi^0$
can be a good criterion for testing the nature of the
$D_{s0}^*(2317)$. A simultaneous study of radiative decays within
the various scenarios is also necessary, as advocated for the light
scalar mesons in Ref.~\cite{sraddec}. To expose the nature of the
$D_{s0}^*(2317)$, experimental efforts are highly appreciated to
improve the quantitative understanding of the $D_{s0}^*(2317)$
decays.

In this paper, we only considered the pseudoscalar $D$--mesons and
the Goldstone bosons. The effects of all the higher states are
incorporated in the LECs. However, the mass difference between $D^*$
and $D$ is only about 140~MeV, which is approximately equal to
$(m_{\Delta}-m_N)/2$. As it is sometimes important to include the
$\Delta(1232)$ in the chiral effective field theory for baryons (for
a recent review, see Ref.~\cite{Bernard:2007zu}), it would be
interesting to check what would happen if we include the vector
charm mesons explicitly in the effective Lagrangian, as in
Ref.~\cite{Lutz:2007sk}. This extension of the scheme will be
investigated in the future.

\section*{Acknowledgments}
We would like to thank A.~Wirzba, M.F.M.~Lutz, M.~Soyeur, and
T.~Gutsche for useful discussions and helpful comments. The work is
partially supported by the Helmholtz Association through funds
provided to the virtual institute ``Spin and strong QCD''(VH-VI-231)
and by the EU Integrated Infrastructure Initiative Hadron Physics
Project under contract number RII3-CT-2004-506078.

\begin{appendix}

\section{Suppression of the $h_6$ terms}
\label{app:h6}

Due to the commutator structure in Eq.~(\ref{eq:L2str}), all  the
$h_6$ terms in amplitudes are proportional to
\beq 
(p_1\cdot p_2)(p_3\cdot p_4) - (p_1\cdot p_4)(p_2\cdot p_3) \ ,
\eeq
 where $p_1$ ($p_2$) and $p_3$ ($p_4$) are the momenta of the
heavy mesons (Goldstone bosons) in the initial and final state,
respectively. Let $v$ denote the velocity of a heavy meson, we
separate the momenta of the heavy mesons into two parts as
\beqa%
p_1 &\!\!=&\!\! M_1 v+k_1, \nonumber\\
p_3 &\!\!=&\!\! M_3 v+k_3,
\eeqa%
where $M_1$ and $M_3$ are the masses of the heavy mesons, and $k_1$
and $k_3$ are small residual momenta which are of order ${\mathcal O}(p)$. Let
$p_3=p_1+\Delta p$ with $\Delta p=\left(M_3-M_1\right)v+k_3-k_1$, we
have \beq (p_1\cdot p_2)(p_3\cdot p_4) - (p_1\cdot p_4)(p_2\cdot p_3) =
\Delta p\cdot\left[\left(p_1\cdot p_2\right)p_4 - \left(p_1\cdot
p_4\right)p_2\right]. \eeq Because $|M_3-M_1|\simeq 100$~MeV at
most, $\Delta p$ should be counted as ${\mathcal O}(p)$. Thus the above
equation should be counted as ${\mathcal O}(p^3)$, and hence is suppressed by
one more order.

\end{appendix}

\end{document}